\documentclass[12pt]{iopart}
\usepackage{iopams}  
\newtheorem{theorem}{Theorem}
\begin{document}

\title[Thermodynamic QED Coherence in Condensed Matter]
{Microscopic Basis of Thermal  Superradiance }

\author{S. Sivasubramanian\dag
\ A. Widom\dag 
\footnote[3]{To whom correspondence should be 
addressed (a.widom@neu.edu)} 
\ and Y.N. Srivastava\dag\ddag }

\address{\dag\ Physics Department, Northeastern University, 
Boston MA 02115, USA}
\address{\ddag\ Dipartimento di Fisica \& INFN, 
Universit\'a di Perugia, Perugia, Italia}

\begin{abstract}
Electromagnetic superradiant field coherence exists  
in a condensed matter system if the electromagnetic 
field oscillators undergo a mean displacement. Transitions 
into thermal states with  ordered superradiant phases have been 
shown to theoretically exist in Dicke-Preparata models. 
The theoretical validity of these models for condensed 
matter has been called into question due to non-relativistic 
diamagnetic terms in the electronic Hamiltonian. The microscopic bases 
of Dicke-Preparata thermal superradiance for realistic macroscopic systems 
are explored in this work. The impossibility of 
diaelectric correlations in condensed matter systems 
(via the Landau-Lifshitz theorem) provides a strong 
theoretical basis for understanding the physical reality of 
condensed matter thermodynamic superradiant phases.  
\end{abstract}
\vspace{.5cm}

\section{Introduction}

In quantum electrodynamic (QED) theory, the 
magnetic field \begin{math} {\bf B}  \end{math} and the Maxwell 
displacement field \begin{math} {\bf D} \end{math} are   
non-commuting quantum fields; i.e. in Gaussian units  
\begin{equation}
\left[D_i({\bf r}),B_j({\bf r}^\prime )\right]=-4\pi i\hbar c 
\epsilon_{ijk}\partial_k \delta ({\bf r}-{\bf r}^\prime ).
\label{canonical}
\end{equation}
The electromagnetic field in a condensed matter system is said to be 
coherent if and only if (on average) 
\begin{math} \overline{\bf B}\ne 0  \end{math}
or \begin{math} \overline{\bf D}\ne 0  \end{math} or both 
situations hold true. In terms of the electromagnetic oscillator modes, 
there is QED coherence \cite{Preparata_95} when one or more  
of the long wavelength field oscillators exhibit a spontaneous non-zero 
mean displacement (even in the absence externally applied fields). 
Ferromagnets and ferroelectrics clearly exhibit 
such spontaneous electromagnetic field ordering on a macroscopic 
scale. Microscopic QED coherence may also occur on the smaller length 
scales of atomic and molecular physics as well as on the length 
scale of mesoscopic coherent domains. For example, atomic or molecular 
magnetic moments clearly exhibit finite magnetic fields 
\begin{math}\overline{\bf B}({\bf r})\ne 0\end{math} in a spatial domain 
surrounding the moments. Dicke models of thermal superradiance have 
been extensively studied [2-23]
%\cite{Hepp_73_1, Hepp_73_2, Wang_73, 
%Carmichel_73, Hioe_73,Vertogem_74,Kudenko_75,Pimentel_75,
%Zaslavsky_75,Gilmore_76_1,Gilmore_76_2,Emelanov_76, Takeno_77, 
%Brito_77,Klemm_77,Guidice_88, Preparata_90_1, Preparata_90_2,
%Preparata_93,Enz_97,Giudice_98,Guidice_00} 
and constitute an important class of ordered electromagnetic 
field coherent states.

The Dicke Hamiltonian\cite{Dicke_54} for a single 
photon oscillator may be written in the form 
\begin{equation}
H_{Dicke}=\frac{1}{2}\left(P^2+\omega_\infty ^2 Q^2 \right)-fQ+
H_{electronic}.
\label{DickeHamiltonian}
\end{equation}
The first term on the right hand side of Eq.(\ref{DickeHamiltonian}) 
represents a single photon oscillator mode. The electronic states are 
described by the Hamiltonian \begin{math} H_{electronic} \end{math}. 
In the Dicke model,  \begin{math} H_{electronic} \end{math} describes 
a set of two (energy) level molecules. The oscillator force operator  
\begin{math} f \end{math} is determined by the electronic dipole moment 
operators of the molecules. The existence of a 
superradiant phase transition has been rigorously proven 
\cite{siva_01_1,siva_01_2} for the strong 
coupling Dicke Hamiltonian in Eq.(\ref{DickeHamiltonian}).
The reason for the superradiant phase transition may be understood 
by computing the shift in the photon oscillator frequency from 
\begin{math} \omega_\infty  \end{math} to 
\begin{math} \omega_0 \end{math} due to the dipole 
interaction. The renormalized frequency \begin{math} \omega_0 \end{math} 
is determined by the oscillator strength sum rule \cite{{siva_01_1}}  
\begin{equation}
\omega_\infty^2 =\omega_0^2 
+\frac{2}{\pi}\int_0^\infty \Gamma (\omega)d\omega 
\ \ {\rm where}\ \ \Gamma (\omega)\ge 0 . 
\label{shift}
\end{equation} 
The oscillator frequency dependent damping function, denoted by   
\begin{math}  \Gamma (\omega) \end{math} and caused by the random force 
\begin{math} f \end{math},
lowers the square of the frequency from  
\begin{math} \omega_\infty ^2 \end{math}
to \begin{math} \omega_0^2 \end{math}. If the damping is sufficiently strong, 
then \begin{math} \omega_0^2 < 0\end{math} and the 
oscillator becomes unstable about the old equilibrium position 
\begin{math} \bar{Q}_{old}=0 \end{math}. A new equilibrium position 
is reached \begin{math} \bar{Q}_{new}\ne 0 \end{math} which describes 
the superradiant ordered state.  

There is little doubt that strong damping can produce a 
superradiant ordered phase if the Dicke Hamiltonian is employed. 
However, the validity of the Dicke model Hamiltonian 
as a description of real condensed matter systems has been 
%\cite{Rzazewski_75,Rzazewski_76_1,Rzazewski_76_2,Bialynicki_79,Gawedzki_81} 
questioned [27-31]. At issue are the diamagnetic terms in a 
more realistic microscopic non-relativistic electronic 
Hamiltonian [32-35].
%\cite{Weisskopf_30,Arecchi_70,Eberly_72,Stenholm_73}. 
With \begin{math}{\bf B}=curl{\bf A} \end{math}, there are 
in the non-relativistic electronic Hamiltonian   
both linear and quadratic terms in \begin{math} {\bf B}  \end{math}. The 
quadratic terms in  \begin{math} {\bf B} \end{math} induce possible 
diamagnetic effects in condensed matter systems.  
Diamagnetism is dominant in some ordered phases of matter 
(e.g. superconductors) but ferromagnetic correlations 
are dominant in some other ordered phases 
of matter (e.g. ferrites). For magnetic field coherence  
\begin{math} \overline{\bf B}\ne 0 \end{math}, 
the frequency shift in Eq.(\ref{shift}) 
is renormalized (in part) by diamagnetic strengths 
independent of the oscillator damping coefficient  
\begin{math} \Gamma (\omega ) \end{math}.  
However, the superradiant ordered phases in the Maxwell displacement  
field \begin{math} \overline{\bf D} \end{math} are {\em not} adversely 
affected by diamagnetic couplings.

The essential physical ideas have been explored by Landau and Lifshitz
\cite{Landau}. The conventional QED Hamiltonian 
contains {\em only linear interaction 
terms} in \begin{math} {\bf D}  \end{math}. 
Landau and Lifshitz prove the following important theorem for  
materials described by a dielectric constant 
\begin{math} \varepsilon  \end{math} 
(wherein \begin{math} \overline{\bf D}= 
\varepsilon \overline{\bf E} \end{math}) 
and a magnetic permeability \begin{math} \mu  \end{math} 
(wherein \begin{math} \overline{\bf B}= \mu \overline{\bf H} \end{math}): 
\begin{theorem}\label{T:LandL}
The magnetic permeability $\mu >0$ can be either 
paramagnetic $\mu >1$ or diamagnetic  $\mu <1$. The dielectric 
constant must be paraelectric $\varepsilon >1$. 
There is no ``diaelectricity'' in condensed matter systems,  
i.e. the range $1>\varepsilon >0$ is strictly forbidden.
\end{theorem}

The above theorem and the generalizations to be discussed below  
provide a strong theoretical basis for physical reality of the 
thermal superradiant phase. Our purpose is to discuss 
the theoretical microscopic bases of Dicke-Preparata models 
as applied to realistic macroscopic condensed matter systems.

In Sec.2 we describe the conventional QED Hamiltonian.  
The electronic degrees of freedom are considered to be 
non-relativistic. A canonical transformation is introduced which 
allows for a precise microscopic definition for the transverse  
Maxwell displacement field 
\begin{math} {\bf D}={\bf E}+4\pi {\bf P} \end{math}. 
In Sec.3 the notion of diabatic and adiabatic changes in  
electrodynamic processes will be defined. Dissipation is 
introduced as a diabatic (i.e. non-adiabatic) process. In 
Sec.4, the statistical thermodynamics of the electric dipole 
moment interactions will be discussed and a generalization 
of the Landau-Lifshitz Theorem \ref{T:LandL} will be proved. 
The diabatic dissipation will be related to the thermodynamic 
dielectric response of the medium. In Sec.5, it will 
be shown how the instability of the thermodynamic dielectric response 
function is the signature of a transition into a superradiant phase.  

\section{QED Hamiltonian}

If we choose a vector potential in the Coulomb gauge, 
\begin{equation} 
{\bf B}=curl{\bf A},\ \ \ div{\bf A}=0, 
\label{CoulombGauge}
\end{equation}
then the QED Hamiltonian of interest in the work which follows 
has the form 
\begin{equation}
H_{QED}=\frac{1}{8\pi }
\int \left(|{\bf E}^\prime |^2+|{\bf B}|^2\right)d^3{\bf r}+H[{\bf A}],
\label{QEDHamiltonian}
\end{equation}
where \begin{math} {\bf E}^\prime \end{math} is the operator which 
denotes the {\em transverse part} of the electric field,  
\begin{equation}
{\bf E}^\prime ({\bf r})=4\pi i\hbar c 
\left(\frac{\delta}{\delta {\bf A}({\bf r})}\right)=4\pi 
i\hbar c\ curl\left(\frac{\delta}{\delta {\bf B}({\bf r})}\right),
\label{TransverseElectric} 
\end{equation}
and \begin{math} H[{\bf A}] \end{math} denotes the Hamiltonian 
(including Coulomb interactions) of the charged particles. The 
Coulomb Hamiltonian when \begin{math} {\bf B}=0   \end{math} 
has the form 
\begin{equation}
H[{\bf A}=0]=H_{Coul}=-\sum_j\frac{\hbar^2}{2m} \Delta_j
-\sum_a \frac{\hbar^2}{2M_a}\Delta_a +U,
\label{CoulombHamiltonian}
\end{equation}
where the total potential energy for the electrons and nuclei in 
the condensed matter system is given by
\begin{equation}
U=e^2\left(\sum_{j<k}\frac{1}{r_{jk}}+\sum_{a<b}\frac{Z_aZ_b}{R_{ab}}
-\sum_{ja}\frac{Z_a}{|{\bf r}_j-{\bf R}_a|}\right).
\label{CoulombPotential}
\end{equation}
For \begin{math} {\bf A}\ne 0  \end{math},
\begin{equation} 
H[{\bf A}]=-\sum_j
\frac{\left(\hbar c\nabla_j-ie{\bf A}_j\right)^2}{2mc^2} -\sum_a 
\frac{\left(\hbar c\nabla_a-iZ_a|e|{\bf A}_a\right)^2}{2M_ac^2}+
H_s+U, 
\label{ElectronicHamiltonian}
\end{equation}
where \begin{math} {\bf A}_j={\bf A}({\bf r}_j) \end{math}, 
\begin{math} {\bf A}_a={\bf A}({\bf R}_a) \end{math} and 
the interaction of the magnetic field with the particle spins 
is given by 
\begin{equation}
H_{s}=-\left(\frac{ge}{2mc}\right)
\sum_j {\bf s}_j\cdot {\bf B}({\bf r}_j)-
\sum_a \left(\frac{g_aZ_a|e|}{2M_ac}\right) 
{\bf S}_a\cdot {\bf B}({\bf R}_a).
\label{SpinInteraction}
\end{equation}

Let us consider a basis in which the Coulomb Hamiltonian is 
diagonal; i.e.
\begin{equation}
H[{\bf A}=0]\psi_j=H_{Coul}\psi_j=W_j[{\bf A}=0]\psi_j 
\ \ {\rm where}\ \ j=0,1,2,\ldots \ .
\label{CoulombBasis}
\end{equation}
In the Coulomb basis, one may define the matrix elements 
\begin{equation}
H_{jk}[{\bf A}]=\left(\psi_k,H[{\bf A}]\psi_j\right)
\label{MatrixHam1}
\end{equation}
so that the Hamiltonian in Eq.(\ref{ElectronicHamiltonian}) 
has the equivalent matrix representation  
\begin{equation}
H[{\bf A}]=\pmatrix{
H_{00}[{\bf A}] & H_{01}[{\bf A}] &  H_{02}[{\bf A}] & \dots \cr 
H_{10}[{\bf A}] & H_{11}[{\bf A}] &  H_{12}[{\bf A}] & \dots \cr
H_{20}[{\bf A}] & H_{21}[{\bf A}] &  H_{22}[{\bf A}] & \dots \cr  
\vdots          & \vdots          & \vdots           & \ddots }.
\label{MatrixHam2}
\end{equation}
In principle, the Hamiltonian can be brought to diagonal 
form by a unitary transformation
\begin{equation}
W[{\bf A}]=U^\dagger [{\bf A}]H[{\bf A}]U[{\bf A}] 
\ \ {\rm where}\ \ U^\dagger [{\bf A}]=U^{-1}[{\bf A}]   
\label{MatrixHam3}
\end{equation}
and
\begin{equation}
W[{\bf A}]=\pmatrix{
W_0[{\bf A}] & 0            & 0            & \dots \cr 
0            & W_1[{\bf A}] & 0            & \dots \cr  
0            & 0            & W_2[{\bf A}] & \dots \cr
\vdots       & \vdots       & \vdots       & \ddots }.
\label{MatrixHam4}
\end{equation}
In virtue of gauge invariance
\begin{equation}
W[{\bf A}]\equiv W[{\bf B}].
\label{Gaugeinv}
\end{equation}
The unitary transformation can be employed to transform the total Hamiltonian 
in Eq.(\ref{QEDHamiltonian}) into the adiabatic representation 
\begin{equation}
{\cal H}=U^\dagger [{\bf A}]H_{QED}U[{\bf A}].
\label{AdiabaticRep1}
\end{equation}
When acting on the electric field in Eq.(\ref{TransverseElectric}), 
the transformation defines the Maxwell displacement field 
\begin{math} {\bf D} \end{math} and  
polarization \begin{math} {\bf P} \end{math} via  
\begin{equation}
U^\dagger [{\bf A}]{\bf E}^\prime U[{\bf A}]={\bf E}={\bf D}-4\pi {\bf P},
\label{NewElectricField}
\end{equation}
where 
\begin{equation}
{\bf D} ({\bf r})=4\pi i\hbar c 
\left(\frac{\delta}{\delta {\bf A}({\bf r})}\right)=4\pi 
i\hbar c\ curl\left(\frac{\delta}{\delta {\bf B}({\bf r})}\right), 
\label{MaxwellDfield}
\end{equation}
and 
\begin{equation}
{\bf P} ({\bf r})= -i\hbar cU^\dagger[{\bf A}] 
\left(\frac{\delta U[{\bf A}]}{\delta {\bf A}({\bf r})}\right)=i\hbar c
\left(\frac{\delta U^\dagger [{\bf A}]}{\delta {\bf A}({\bf r})}\right)
U[{\bf A}]. 
\label{Polarization}
\end{equation}
Eqs.(\ref{QEDHamiltonian}), (\ref{TransverseElectric}) and  
(\ref{AdiabaticRep1}) - (\ref{Polarization}) 
imply the adiabatic Hamiltonian representation 
\begin{equation}
{\cal H}=\frac{1}{8\pi }
\int \left(|{\bf D}-4\pi {\bf P} |^2
+|{\bf B}|^2\right)d^3{\bf r}+W[{\bf A}],
\label{AdiabaticRep2}
\end{equation}
which may be conveniently written as 
\begin{equation}
{\cal H}=\frac{1}{8\pi }\int |{\bf D}|^2d^3{\bf r}
-\int {\bf D}\cdot {\bf P}d^3{\bf r}+H^\prime [{\bf A}].
\label{AdiabaticRep3}
\end{equation}
In Eq.(\ref{AdiabaticRep3}),  
\begin{equation}
H^\prime [{\bf A}]=\frac{1}{8\pi }\int |{\bf B}|^2d^3{\bf r}
+W[{\bf A}]+2\pi \int |{\bf P}|^2d^3{\bf r}. 
\label{AdiabaticRep4}
\end{equation}
Finally, the transverse current operator in the adiabatic 
representation of Eqs.(\ref{MatrixHam3}) and (\ref{MatrixHam4}) 
is given by 
\begin{equation}
{\bf J}({\bf r})=-cU^\dagger [{\bf A}]\frac{\delta H[{\bf A}]}
{\delta {\bf A}({\bf r})}U[{\bf A}].
\label{Current1}
\end{equation}
The notion of ``diabatic damping'' is associated with a decomposition 
of the current Eq.(\ref{Current1}) into adiabatic and non-adiabatic 
parts.

\section{Damping and Electrical Conductivity}

For the unitary transformation in Eq.(\ref{MatrixHam3}) we have 
the differential identity 
\begin{equation}
U^\dagger \frac{\delta H}{\delta {\bf A}}U
=\frac{\delta}{\delta {\bf A}}
\left(U^\dagger HU\right)-U^\dagger HU U^\dagger 
\frac{\delta U}{\delta {\bf A}}
-\frac{\delta U^\dagger}{\delta {\bf A}}UU^\dagger HU,  
\label{Differential1}
\end{equation}
which together with Eqs.(\ref{Polarization}) and 
(\ref{Current1}) yields the current decomposition 
\begin{equation}
{\bf J}({\bf r})=-c\frac{\delta W[{\bf A}]}{\delta {\bf A}({\bf r})}
+\frac{i}{\hbar }\left[W[{\bf A}],{\bf P}({\bf r})\right].
\label{Current2}
\end{equation} 

One may employ the gauge invariant 
\begin{math} {\bf B}=curl{\bf A} \end{math} in the form 
[see Eq.(\ref{Gaugeinv})] of an adiabatic current 
\begin{equation}
{\bf J}_{adiabatic}({\bf r})
=-c\frac{\delta W[{\bf A}]}{\delta {\bf A}({\bf r})}=
-c\ curl \frac{\delta W[{\bf B}]}{\delta {\bf B}({\bf r})}=
c\ curl{\bf M}.
\label{Magnetization1}
\end{equation}
The gauge invariant adiabatic magnetization has a form which 
follows from Eqs.(\ref{MatrixHam4}) and (\ref{Magnetization1}); i.e. 
\begin{equation}
{\bf M}=-\pmatrix{
\delta W_0[{\bf B}]/\delta {\bf B} 
& 0            & 0            & \dots \cr 
0            & \delta W_1[{\bf B}]/\delta {\bf B}  
& 0            & \dots \cr  
0            & 0            & \delta W_2[{\bf B}]/\delta {\bf B}  
& \dots \cr
\vdots       & \vdots       & \vdots       & \ddots }.
\label{Magnetization2}
\end{equation}
The dissipative (or ``diabatic'') part of the current operator is given 
by 
\begin{equation}
{\bf J}_d({\bf r})=\frac{i}{\hbar }\left[W,{\bf P}({\bf r})\right]
=\dot{\bf P}({\bf r}).
\label{Current3}
\end{equation}
Eq.(\ref{Current2}) then takes the conventional form having both 
magnetization and polarization parts; i.e. 
\begin{equation}
{\bf J}({\bf r})=c\ curl{\bf M}({\bf r})+\dot{\bf P}({\bf r}).
\label{Current4}
\end{equation}
Although Eq.(\ref{Current4}) is well known in classical electrodynamics, 
we have given the proof from a fully quantum electrodynamic viewpoint. 

The diabatic part of the current in Eq.(\ref{Current3}) describes 
the dissipation via the non-local transverse electrical conductivity 
tensor  
\begin{math} 
\sigma_{ij}({\bf r},{\bf r}^\prime ,\omega +i0^+ ) 
\end{math}. The microscopic expression for the transverse 
conductivity is determined by the fluctuation dissipation theorem   
\begin{equation}
\sigma_{ij}({\bf r},{\bf r}^\prime ,\zeta )=
\frac{1}{\hbar }\int_0^\beta \int_0^\infty e^{i\zeta t}
\left<J_{d,j}({\bf r}^\prime ,-i\lambda )J_{d,i}({\bf r},t)\right>
dt d\lambda .
\label{conduct1}
\end{equation}
Note: (i) The complex frequency obeys 
\begin{math} \Im m\ \zeta >0 \end{math}. (ii) The thermal average 
\begin{math} \left< \dots \right> \end{math} is over the charged 
particle degrees of freedom. (iii) The time variation of operators 
in Eq.(\ref{conduct1}) employs  
the Hamiltonian \begin{math} H^\prime [{\bf A}] \end{math} in  
Eq.(\ref{AdiabaticRep4}). (iv) Finally, 
\begin{equation}
\beta =\left(\frac{\hbar }{k_B T}\right).
\label{Temperature}
\end{equation}

For the transverse dielectric properties of charged particles, 
one employs the electric susceptibility 
\begin{equation}
\chi_{ij}({\bf r},{\bf r}^\prime ,\zeta )=\frac{i}{\hbar }
\int_0^\infty e^{i\zeta t}
\left<\left[P_i({\bf r},t),P_j({\bf r}^\prime ,0)\right]\right>dt.
\label{chi1}
\end{equation}
Integrating Eq.(\ref{chi1}) by parts yields 
\begin{equation}
-i\zeta \chi_{ij}({\bf r},{\bf r}^\prime ,\zeta )=
h_{ij}({\bf r},{\bf r}^\prime )+\frac{i}{\hbar }
\int_0^\infty e^{i\zeta t}
\left<\left[\dot{P}_i({\bf r},t),
P_j({\bf r}^\prime ,0)\right]\right>dt,
\label{chi2}
\end{equation}
where the equal time commutator contribution is given by 
\begin{equation}
h_{ij}({\bf r},{\bf r}^\prime )=\frac{i}{\hbar }
\left<\left[P_i({\bf r}),P_j({\bf r}^\prime )\right]\right>.
\label{chi3}
\end{equation}
Employing the Kubo-Martin-Schwinger [37-40]
%\cite{Kubo_57,Martin_59,Kadanoff_63,Martin_68} 
condition  
\begin{equation}
i\left<\left[\dot{P}_i({\bf r},t),P_j({\bf r}^\prime ,0)\right]\right>=
\int_0^\beta \left<\dot{P}_j({\bf r}^\prime ,-i\lambda) 
\dot{P}_i({\bf r},t)\right>d\lambda 
\label{chi4}
\end{equation}
and Eq.(\ref{Current3}) in Eq.(\ref{conduct1}) yields a simple 
relationship between the conductivity and the susceptibility   
\begin{equation}
-i\zeta \chi_{ij}({\bf r},{\bf r}^\prime ,\zeta )
=\sigma_{ij}({\bf r},{\bf r}^\prime ,\zeta )
+h_{ij}({\bf r},{\bf r}^\prime ).
\label{chi5}
\end{equation}
The dissipative part of the conductivity is determined by 
\begin{equation}
{\Re e }\{\sigma_{ij}({\bf r},{\bf r}^\prime ,\omega +i0^+)\}
=\omega {\Im m}\{\chi_{ij}({\bf r},{\bf r}^\prime ,\omega +i0^+)\}.
\label{chi6}
\end{equation}
The quantum mechanical hall conductivity contribution 
\begin{math} h_{ij}({\bf r},{\bf r}^\prime ) \end{math} from 
the transverse polarization is determined by the commutation 
relation in Eq.(\ref{chi3}). 

Note that the response functions discussed above are defined 
with respect to an applied magnetic field 
\begin{math} {\bf B}=curl{\bf A} \end{math}. 
Magneto-conductivity in \begin{math} \sigma_{ij}  \end{math} and 
the Faraday effect in \begin{math} \chi_{ij} \end{math} are implicitly 
included in the above considerations. For superradiance described 
by a coherent field \begin{math} \overline{\bf D}\ne 0 \end{math} 
one averages over magnetic field fluctuations. The mean 
magnetic field (on macroscopic length scales)  
obeys \begin{math} \overline{\bf B}= 0 \end{math}. 
Let us now consider the details of the magnetic field 
fluctuations. 

\section{Statistical Thermodynamics}

The quantum electrodynamic free energy \begin{math} F \end{math} 
of a condensed matter system can be written as 
\begin{equation}
{\cal G}=-k_BT\ln \left\{Tr\ e^{-{\cal H}/k_BT} \right\}
\label{Thermo1}
\end{equation}
where \begin{math} {\cal H} \end{math} is given in 
Eq.(\ref{AdiabaticRep3}). The complete trace over quantum states includes 
electromagnetic field degrees of freedom as well as charged particle degrees 
of freedom. Thus 
\begin{equation}
Tr\ e^{-{\cal H}/k_BT} =Tr_{({\bf D},{\bf A})}
\left(Tr_{(Charged)}\ e^{-{\cal H}/k_BT}\right). 
\label{Thermo2}
\end{equation}
If one treats the electromagnetic field trace in the quasi-classical 
limit of a functional integral  
\begin{equation}
Tr_{({\bf D},{\bf A})}(\ldots )\to \int \prod_{\bf r} 
\left(\frac{{\cal D}{\bf D}({\bf r}){\cal D}{\bf A}({\bf r})}
{8\pi \hbar c} \right)(\ldots ), 
\label{Thermo3}
\end{equation}
then Eqs.(\ref{AdiabaticRep3}) and (\ref{Thermo1})-(\ref{Thermo3}) 
imply the functional representation 
\begin{equation}
\exp \left(-\frac{\cal G}{k_BT}\right)
=\int \exp \left(-\frac{{\cal F}_{tot}[{\bf D}]}{k_BT}\right)
\prod_{\bf r}\left({\cal D}{\bf D}({\bf r})\right).
\label{Thermo4}
\end{equation}
The total free energy functional of the Maxwell displacement 
field obeys 
\begin{equation}
{\cal F}_{tot}[{\bf D}]=
\frac{1}{8\pi }\int |{\bf D}|^2 d^3{\bf r}+{\cal F}[{\bf D}],
\label{Thermo5}
\end{equation}
where 
\begin{equation}
\exp \left(-\frac{{\cal F}[{\bf D}]}{k_BT}\right)=Tr_{(Charged)}\int 
\exp \left(-\frac{H^{\prime \prime}[{\bf A},{\bf D}]}{k_BT}\right)
\prod_{\bf r}
\left(\frac{{\cal D}{\bf A}({\bf r})}{8\pi \hbar c}\right) 
\label{Thermo6}
\end{equation}
and 
\begin{equation}
H^{\prime \prime}[{\bf A},{\bf D}]=H^\prime [{\bf A}]
-\int {\bf P}({\bf r},[{\bf A}])\cdot{\bf D}({\bf r})d^3{\bf r}.
\label{Thermo7}
\end{equation}

The coupling into the Maxwell displacement field in the Hamiltonian 
Eq.({\ref{Thermo7}}) is linear. Eqs.({\ref{Thermo6}}) and 
({\ref{Thermo7}}) imply that the free energy 
\begin{math} {\cal F}[{\bf D}] \end{math} is a convex upward functional.
For the second functional derivative
\begin{equation}
\tilde{\chi }_{ij}({\bf r},{\bf r}^\prime ,[{\bf D}])=
-\left(\frac{\delta ^2  {\cal F}[{\bf D}]}
{\delta D_i({\bf r})\delta D_j({\bf r}^\prime )}\right)
\label{Thermo8}
\end{equation}
we have the following:
\begin{theorem}\label{T:LandLandS1}
The spectrum of the zero frequency susceptibility  
$\{\tilde{\chi }_\lambda [{\bf D}]\}$ 
defined by the eigenvalue equation 
$$
\sum_j \int \tilde{\chi }_{ij}({\bf r},{\bf r}^\prime ,[{\bf D}])
\xi^j_\lambda ({\bf r}^\prime )d^3{\bf r}^\prime =
\tilde{\chi }_\lambda [{\bf D}]\xi^i_\lambda ({\bf r})
$$  
obey $\tilde{\chi }_\lambda [{\bf D}]\ge 0$ for all $\lambda $.
\end{theorem}
In order to prove a generalization of the Landau and Lifshitz 
Theorem \ref{T:LandL}, one must investigate the second functional 
derivatives 
\begin{eqnarray}
\eta_{ij}({\bf r},{\bf r}^\prime ,[{\bf D}])&=&4\pi 
\left(\frac{\delta ^2  {\cal F}_{tot}[{\bf D}]}
{\delta D_i({\bf r})\delta D_j({\bf r}^\prime )}\right)
\nonumber \\ 
&=&\Delta_{ij}({\bf r}-{\bf r}^\prime )
-4\pi  \tilde{\chi }_{ij}({\bf r},{\bf r}^\prime ,[{\bf D}]),
\label{Thermo9}
\end{eqnarray}
where Eq.(\ref{Thermo5}) has been invoked and the transverse 
delta function is defined as  
\begin{eqnarray}
\Delta_{ij}({\bf r})&=&\int e^{i{\bf k \cdot r}}
\left(\delta_{ij}-\frac{k_ik_j}{|{\bf k}|^2}\right)
\frac{d^3{\bf k}}{(2\pi )^3}\ , \nonumber \\
&=& \left(\frac{2}{3}\right)\delta_{ij}\delta ({\bf r})
+\left(\frac{1}{4\pi }\right)
\left(\frac{3r_i r_j-r^2\delta_{ij}}{r^5}\right). 
\label{Thermo10}
\end{eqnarray}
The spectrum of the zero frequency transverse dielectric 
constants    
\begin{math} \{\varepsilon_\lambda [{\bf D}]\} \end{math} 
is defined by the eigenvalue equation 
\begin{equation}
\sum_j \int \eta_{ij}({\bf r},{\bf r}^\prime ,[{\bf D}])
\xi^j_\lambda ({\bf r}^\prime )d^3{\bf r}^\prime =
\left(\frac{1}{\varepsilon_\lambda [{\bf D}]}\right)
\xi^i_\lambda ({\bf r}),
\label{Thermo11}
\end{equation}
and obeys 
\begin{equation}
\varepsilon_\lambda [{\bf D}]
=\left[\frac{1}{1-4\pi \tilde{\chi }_\lambda [{\bf D}]}\right].
\label{Thermo12}
\end{equation}

The thermodynamic stability condition for the free energy  
\begin{math} {\cal F}_{tot}[{\bf D}]  \end{math} can be summarized by the 
following:
\begin{theorem}\label{T:LandLandS2}
Thermodynamic stability for a Maxwell displacement field ${\bf D}$ 
requires that $\varepsilon_\lambda [{\bf D}]>0$ for all $\lambda $.
\end{theorem}
{\em Proof:} For a Maxwell field \begin{math} {\bf D} \end{math} 
to represent thermal equilibrium, one expects the free energy 
\begin{math}  {\cal F}_{tot}[{\bf D}] \end{math} to be at a minimum. 
The second (functional) derivative conditions for achieving a 
thermodynamic free energy minimum are those stated in the theorem 
in virtue of Eqs.(\ref{Thermo9}) and (\ref{Thermo11}).

The central result of this section is the {\em generalization} of 
the Landau-Lifshitz Theorem \ref{T:LandL}:  
\begin{theorem}\label{T:LandLandS3}
Thermodynamic stability for a Maxwell displacement field ${\bf D}$ 
requires the paraelectric inequality $\varepsilon_\lambda [{\bf D}]>1$ 
for all $\lambda $. Diaelectric behavior 
$(0<\varepsilon_\lambda [{\bf D}]<1)$ is strictly forbidden.
\end{theorem}
{\em Proof:} Theorem \ref{T:LandLandS3} follows directly from 
Theorem \ref{T:LandLandS1}, Eq.(\ref{Thermo12}) and 
Theorem \ref{T:LandLandS2}. 

We now reconsider the dissipative properties of a 
condensed matter system {\em after} coherent averaging over magnetic 
field \begin{math} {\bf B}=curl{\bf A} \end{math} fluctuations. 
We employ the functional averaging measure 
\begin{math} 
\int (\ldots )\prod_{\bf r}({\cal D}{\bf A}/8\pi \hbar c) 
\end{math}
as in Eqs.(\ref{Thermo5}) and (\ref{Thermo6}). For example, 
the new susceptibility 
\begin{equation}
\tilde{\chi }_{ij}({\bf r},{\bf r}^\prime ,\zeta ,[{\bf D}] )
=\frac{i}{\hbar }\int_0^\infty e^{i\zeta t}
\left<\left<\left[P_i({\bf r},t),P_j({\bf r}^\prime ,0)\right]
\right>\right>dt,
\label{Thermo13}
\end{equation}
where the complete ``double averaging''  
\begin{math}\left<\left< \ldots  \right>\right> \end{math} 
is over both charged particle motions and magnetic field 
fluctuations with the Maxwell displacement field 
\begin{math} {\bf D} \end{math} held fixed. The time dependent 
operators in Eq.(\ref{Thermo13}) are with respect to the Hamiltonian 
\begin{math} H^{\prime \prime}[{\bf A},{\bf D}] \end{math} defined 
in Eq.(\ref{Thermo7}). The dissipative motions of the condensed 
matter system are then described by 
\begin{equation}
{\Re e }\{\tilde{\sigma}_{ij}
({\bf r},{\bf r}^\prime ,\omega +i0^+,[{\bf D}])\}=\omega {\Im m}
\{\tilde{\chi}_{ij}({\bf r},{\bf r}^\prime ,\omega +i0^+,[{\bf D}])\}.
\label{Thermo14}
\end{equation}
Finally, the static susceptibility in Eq.(\ref{Thermo8}) is the zero 
frequency limit
\begin{equation}
\tilde{\chi }_{ij}({\bf r},{\bf r}^\prime ,[{\bf D}])\equiv 
\lim_{\omega \to 0}{\Re e }
\{\tilde{\chi}_{ij}({\bf r},{\bf r}^\prime ,\omega +i0^+,[{\bf D}])\}.
\label{Thermo15}
\end{equation}
The thermodynamic stability test for a coherent electromagnetic 
superradiant state follows from Theorem \ref{T:LandLandS1},
Eq.(\ref{Thermo12}) and Theorem \ref{T:LandLandS3}. 

\section{Thermodynamic Stability}

In the absence of external electric fields, the stable values of 
\begin{math} {\bf D} \end{math} are such as to 
minimize \cite{Landau} the total free energy 
\begin{math} {\cal F}_{tot}[{\bf D}]\end{math}. The free energy 
minimization condition implies a vanishing thermal 
electric field;  
\begin{equation}
{\bf E}({\bf r})=4\pi \left(\frac{\delta {\cal F}_{tot}[{\bf D}]}
{\delta {\bf D}({\bf r})}\right)= {\bf D}({\bf r})-
4\pi\left<\left< {\bf P}({\bf r})\right>\right>=0.
\label{TS1}
\end{equation}
The first derivative Eq.(\ref{TS1}) will in general have more than one 
possible solution with a non-zero Maxwell displacement field 
\begin{math} [{\bf D}] \end{math}. The multiple solutions correspond 
to differing possibilities for superradiant coherent domains.
A necessary condition for a true free energy minimum 
has been proved in Theorem \ref{T:LandLandS3}. Employing 
Eq.(\ref{Thermo12}) we find the stability condition 
\begin{equation}
0<4\pi \tilde{\chi }_\lambda [{\bf D}]<1.
\label{TS2}
\end{equation}
From (i)Eq.(\ref{Thermo13}), (ii) the resulting dispersion relation  
\begin{eqnarray}
\tilde{\chi }_{ij}({\bf r},{\bf r}^\prime ,\zeta ,[{\bf D}] )
&=&\frac{2}{\pi }\int_0^\infty 
\frac{{\Im m}\tilde{\chi }_{ij}
({\bf r},{\bf r}^\prime ,\omega +i0^+ ,[{\bf D}] )d\omega}
{\omega^2-\zeta ^2} \nonumber \\ 
\tilde{\chi }_{ij}({\bf r},{\bf r}^\prime ,[{\bf D}])
&\equiv & \lim_{\zeta \to 0}
\tilde{\chi }_{ij}({\bf r},{\bf r}^\prime ,\zeta ,[{\bf D}])
\nonumber \\
&=&\frac{2}{\pi }\int_0^\infty {\Im m}\tilde{\chi }_{ij}
({\bf r},{\bf r}^\prime ,\omega +i0^+ ,[{\bf D}]) 
\frac{d\omega }{\omega },
\label{TS3}
\end{eqnarray}
(iii) Theorem{\ref{T:LandLandS1}} and (iv) the stability 
Eq.(\ref{TS2}), it follows that strong dissipation, i.e. substantial
\begin{math} 
\{{\Im m}\tilde{\chi }(\omega +i0^+)/\omega \}
\end{math}, 
tends to yield thermodynamic instabilities. 

Symmetry under parity transformations yield the solution 
\begin{math} {\bf D}=0 \end{math}
corresponding to a ``normal'' phase. If 
\begin{math} 4\pi \tilde{\chi }_\lambda [{\bf D}=0]>1 \end{math} 
for some \begin{math} \lambda \end{math}, then the normal 
phase is unstable. The true free energy minimum will arise 
for coherent superradiant domains with  
\begin{math} [{\bf D}\ne 0] \end{math}.  There may be many possible 
superradiant domain configurations as discussed above.
Domain walls  and/or normal phase regions may separate 
the superradiant domains wherein 
\begin{math} 
{\bf D}=4\pi \left<\left< {\bf P}({\bf r})\right>\right>\ne 0  
\end{math}.

\section{Conclusions}

We have discussed a generalization of the theorem by Landau and 
Lifshitz whereby \begin{math} \varepsilon \ge 1 \end{math} for 
materials described by \begin{math} {\bf D}=\varepsilon {\bf E} \end{math}.  
The theorem and its generalization is {\em independent} of magnetic 
permeability \begin{math} \mu \end{math} when the material also obeys 
\begin{math} {\bf B}=\mu {\bf H} \end{math}. Magnetic properties may 
be either paramagnetic or diamagnetic, but diaelectric properties 
of matter are ruled out by quantum statistical mechanical stability 
considerations. 

The stability criteria are crucial for the understanding of stable 
superradiant domains. In particular, for all of the possible 
eigen modes of the dielectric susceptibility, the generalized 
Landau-Lifshitz stability condition 
\begin{math} \varepsilon_\lambda[{\bf D}] \ge 1 \end{math} 
is crucial for testing whether (or not) a superradiant domain 
is thermodynamically stable. Contrary to what has some time ago 
appeared in the literature \cite{Rzazewski_76_2} the dielectric 
stability conditions are independent of possible diamagnetism. 
No approximations to the free energy functional 
\begin{math} {\cal F}_{tot}[{\bf D}]  \end{math} have been required 
to reach important general conclusions, however further approximations 
{\em are required} for concrete computation.

\end{document}